\documentclass[english,aps, pre,reprint]{revtex4-1}
\usepackage[latin9]{inputenc}
\usepackage{color}
\usepackage{amsmath}
\usepackage{amssymb}
\usepackage{graphicx}
\usepackage{esint}
\makeatletter
 \@ifundefined{textcolor}{}
 {%
   \definecolor{BLACK}{gray}{0}
   \definecolor{WHITE}{gray}{1}
   \definecolor{RED}{rgb}{1,0,0}
   \definecolor{GREEN}{rgb}{0,1,0}
   \definecolor{BLUE}{rgb}{0,0,1}
   \definecolor{CYAN}{cmyk}{1,0,0,0}
   \definecolor{MAGENTA}{cmyk}{0,1,0,0}
   \definecolor{YELLOW}{cmyk}{0,0,1,0}
 }

\usepackage[english]{babel}

\makeatother

\usepackage{babel}
\begin{document}

\title{Liquid to solid nucleation via onion structure droplets}
\author{Kipton Barros}
\affiliation{Theoretical Division and CNLS, Los Alamos National Laboratory, Los Alamos, New Mexico 87544}
\author{W. Klein}
\affiliation{Physics Department, Boston University, Boston, Massachusetts 02215}

\begin{abstract}
We study homogeneous nucleation from a deeply quenched metastable liquid to a spatially modulated phase. We find, for a general class of density functional theories, that the universally favored nucleating droplet in dimensions $d \geq 3$ is spherically symmetric with radial modulations resembling the layers of an onion. The existence of this droplet has important implications for systems with effective long-range interactions, and potentially applies to polymers, plasmas, and metals.
\end{abstract}

% crystallization in liquid-solid (64.70.dg)
% nucleation in phase transitions (64.60.Q-)
% theory and models of crystal growth (81.10.Aj)
% \pacs{64.70.dg, 64.60.Q-, 81.10.Aj}

\maketitle

\global\long\def\mathd{\,\mathrm{d}}

\section{Introduction}

Modulated pattern formation occurs in a wide variety of systems.
Examples include cholesteric liquid crystals~\cite{Brazovskii75},
hydronamic instabilities~\cite{Swift77}, superconducting vortices~\cite{Pardo98},
block copolymers~\cite{Harrison00}, as well as many others~\cite{Seul95}.
In recent years, there has been considerable renewed interest is the
\emph{dynamical} process of a first-order phase transition from a
uniform (liquid) phase to a modulated (solid) phase following a quench~\cite{Teeffelen08,Tegze09,Backofen10,Toth10,Toth11,Toth12,Granasy11}.
If the supercooled liquid phase is long-lived, rare metastable equilibrium
fluctuations locally initiate the phase transformation
dynamics. The nucleation barrier is the free energy cost of the \emph{nucleating droplet}, a saddle point solution of the effective free energy functional. Gibbs'  theory of homogeneous nucleation, valid near
liquid-solid coexistence, postulated nucleating droplets as isolated regions of the stable phase separated from the metastable phase by a sharp interface~\cite{Gibbs48}. Subsequent refinements to this theory
apply to deeper quenches within the metastable phase~\cite{Cahn59,Langer67,Gunton99}.
Near the \emph{spinodal} (the limit at which the metastable phase becomes unstable) nucleating droplets
differ dramatically from the stable phase~\cite{Unger84}, a phenomenon
sometimes called the ``Ostwald step rule''~\cite{Ostwald97}.

In liquid to solid nucleation, even the lattice symmetry of
the nucleating droplet may differ from the stable solid phase. Using
classical density functional theory (DFT)~\cite{Oxtoby91}, Alexander
and McTague predicted that a solid phase with a small density difference
$\psi(\mathbf{r})$ from the liquid phase would have hexagonal or
bcc symmetry~\cite{Alexander78,Chaikin95-bcc}. It was subsequently observed that such solid phases, being derived from a free energy functional of order $\psi^{3}$, are actually unstable~\cite{Klein01}.
The Alexander-McTague argument \emph{may}, however, be used to characterize unstable
nucleating droplets~\cite{Klein86}. That is, nucleating droplets near the spinodal are expected to have hexagonal or bcc symmetry.

In this work, we demonstrate a \emph{new} symmetry consistent with a $\psi^{3}$ free energy functional and not considered by Alexander and McTague: a spherically symmetric droplet with
radial modulations reminiscent of layers in an onion. We show that for deep
quenches near the spinodal, where a $\psi^{3}$ theory is valid, this onion structure droplet is in fact
universally favored for a broad class of DFT models. Our
argument applies when the metastable liquid phase is sufficiently long-lived, so that nucleation is well described by a saddle point approximation. For deep quenches, this approximation may be valid when fluctuations are damped by averaging over an effective long interaction range, \emph{e.g.}, the radius of gyration for block copolymer melts~\cite{Fraaije93} or the effective screening length for plasmas~\cite{Hammerberg94}.

There is much related work on liquid to solid nucleation. Our theory applies to
an especially simple DFT, the phase field crystal model
(PFC)~\cite{Elder04}, which has been the subject of much recent interest.
Numerical calculation of PFC saddle points, which represent nucleating droplets,
has been performed in two and three dimensions~\cite{Backofen10,Toth10}.
In dynamical PFC studies, the metastable liquid was found to contain
an amorphous precursor, which promotes the growth of bcc nuclei~\cite{Toth10,Toth11}.
These and other PFC results are reviewed in Ref.~\onlinecite{Granasy11}. Several studies suggest that nucleation is affected by the presence of a nearby (pseudo) spinodal~\cite{Yang88,Yang90,Talanquer98}.
In molecular dynamics (MD) of Lennard-Jones systems quenched not too close to the spinodal, evidence
was found for nucleating droplets with an fcc core and a high degree of bcc ordering at the interface~\cite{Wolde95}. Consistent results were found in DFT
studies of Lennard-Jones~\cite{Shen96a}. Subsequently, droplets of various other structures have been observed~\cite{Wolde99,Cherne04,Wang07}, indicating that nucleation phenomena in these models of atomistic crystallization is not fully controlled by proximity to a spinodal.

The rest of this paper is structured as follows. In Sec.~\ref{sec:prelim}
we define the DFT model and reduce it to a $\psi^{3}$ theory when
liquid phase fluctuations are small. In Sec.~\ref{sec:ls_droplets}
we review the theory of lattice structure nucleating droplets derived
by Klein and Leyvraz~\cite{Klein86}.  In Sec.~\ref{sec:os_droplets} we demonstrate
the existence of an onion structure droplet with qualitatively new scaling
behavior. In Sec.~\ref{sec:energies} we calculate the free energy scaling
of the lattice and onion structure droplet types and
conclude that onion structure droplets are favored in $d\geq3$ dimensions,
independent of the details of the model. Our result is valid for
systems with effective long-range interactions, and for quenches near
the spinodal.

\section{Preliminaries\label{sec:prelim}}

We assume that the effective free energy of the density field $\rho(\mathbf{x})$
has the form,
\begin{equation}
F[\rho]=\!\int\left[\frac{1}{2}\rho(\mathbf{x})(C*\rho)(\mathbf{x})+f(\rho(\mathbf{x}))-h\rho(\mathbf{x})\right]\mathd^{d}\mathbf{x},\label{eq:free_rho}
\end{equation}
where the convolution operation is denoted as
\begin{equation}
(C*\rho)(\mathbf{x})=\!\int C(\mathbf{x'}-\mathbf{x})\rho(\mathbf{x'})\mathd^{3}\mathbf{x'}.
\end{equation}
In (metastable) equilibrium, fluctuations are Boltzmann distributed,
\begin{equation}
P[\rho]\propto\exp(-F[\rho]/k_{B}T),
\end{equation}
with $k_{B}$ the Boltzmann constant and $T$ the temperature. The
term $f(\rho(\mathbf{x}))$ represents a local free energy cost and
the quantity $h$ represents the chemical potential. We choose a symmetric
Kac potential~\cite{Kac63} with characteristic length scale $R$,
\begin{equation}
C(\mathbf{x})=R^{-d}\Lambda(|\mathbf{x}|/R).\label{eq:kac}
\end{equation}

The free energy functional in Eq.~\eqref{eq:free_rho} can be used to model a variety of pattern
forming systems~\cite{Seul95}. In classical DFT of atomistic systems,
$C(\mathbf{x})$ is interpreted as a direct correlation function~\cite{Ramakrishnan79}.
Alternatively, this free energy can be rigorously justified for the \emph{clump model}~\cite{Grewe77a,Grewe77},
in which a system of particles interact via the long-range \emph{repulsive}
step function potential
\begin{equation}
\Lambda(|\mathbf{x}|) =\Theta(R-|\mathbf{x}|)\label{eq:step1}
\end{equation}
with $R\rightarrow\infty$. Perhaps counterintuitively, the particles minimize potential energy by forming ``clumps" of characteristic separation distance $R$~\cite{Klein94}. The temperature controls the magnitude of local entropic free energy,
\begin{equation}
f(\rho(\mathbf{x})) =T\rho(\mathbf{x})\ln\rho(\mathbf{x}),\label{eq:step2}
\end{equation}
and drives a phase transition between a liquid phase of uniform $\rho(\mathbf x) = \rho_0$ and a clump phase of modulated $\rho(\mathbf x)$.

It is convenient to work with dimensionless lengths,
\begin{equation}
\mathbf{r}=\mathbf{x}/R,\quad\phi(\mathbf{r})=\rho(\mathbf{x}).
\end{equation}
The parameter $R$ damps fluctuations by setting an overall
energy scale,
\begin{equation}
F[\phi]=R^{d}\!\int\!\left[\frac{1}{2}\phi(\mathbf{r})(\Lambda*\phi)(\mathbf{r})+f(\phi(\mathbf{r}))-h\phi(\mathbf{r})\right]\!\mathd^{d}\mathbf{r}.\label{eq:free_phi}
\end{equation}
Equilibrium states are solutions to the Euler-Lagrange equation
\begin{equation}
\frac{\delta F}{\delta\phi(\mathbf{r})}=(\Lambda*\phi)(\mathbf{r})+f'(\phi(\mathbf{r}))-h=0,\label{eq:EL_phi}
\end{equation}
where $f'(\phi(\mathbf{r}))$ denotes the ordinary
derivative $\mathd f(y)/\mathd y$ evaluated at $y=\phi(\mathbf{r})$.
We assume that one such solution is a \emph{metastable} liquid phase
of uniform density, $\phi(\mathbf{r})=\phi_{0}$. When $R$ is large,
we expect fluctuations about the liquid phase, $\psi(\mathbf{r})=\phi(\mathbf{r})-\phi_{0}$,
to be small, and we expand $F$ in powers of $\psi(\mathbf{r})$ with 
\begin{align}
F_{0}= & R^{d}\!\int\left[\frac{1}{2}\phi_{0}(\Lambda*\phi_{0})+f(\phi_{0})-h\phi_{0}\right]\mathd^{d}\mathbf{r}\\
F_{1}= & R^{d}\!\int\left[\Lambda*\phi_{0}+f'(\phi_{0})-h\right]\psi(\mathbf{r})\mathd^{d}\mathbf{r}\\
F_{2}= & R^{d}\!\int\left[\frac{1}{2}\psi(\mathbf{r})(\Lambda*\psi)(\mathbf{r})+\frac{1}{2!}f''(\phi_{0})\psi^{2}(\mathbf{r})\right]\mathd^{d}\mathbf{r}\\
F_{3}= & R^{d}\!\int\frac{1}{3!}f'''(\phi_{0})\psi^{3}(\mathbf{r})\mathd^{d}\mathbf{r}.
\end{align}
$F_{0}$ is independent of $\psi(\mathbf{r})$ and can be ignored.
$F_{1}$ is zero because $\phi_{0}$ satisfies the Euler-Lagrange
equation. The two relevant leading order terms are
\begin{equation}
F=F_{2}+F_{3}+\mathcal{O}(\psi^{4}).\label{eq:free_psi}
\end{equation}
Analysis proceeds most easily in Fourier space,
\begin{align}
F_{2} & =\frac{1}{2}R^{d}\!\int A(|\mathbf{k}|)|\psi(\mathbf{k})|^{2}\mathd^{d}\mathbf{k}\label{eq:f2}\\
F_{3} & =-\frac{b}{3}R^{d}\!\int\psi(\mathbf{k})\psi(\mathbf{k}')\psi(-\mathbf{k}-\mathbf{k}')\mathd^{d}\mathbf{k}\mathd^{d}\mathbf{k}',\label{eq:f3}
\end{align}
where
\begin{align}
\psi(\mathbf{k}) & =\!\int e^{i\mathbf{k}\cdot\mathbf{r}}\psi(\mathbf{r})\mathd^{d}\mathbf{r}\\
A(|\mathbf{k}|) & =\frac{1}{(2\pi)^{d}}\left[\int e^{i\mathbf{k}\cdot\mathbf{r}}\Lambda(|\mathbf{r}|)\mathd^{d}\mathbf{r}+f''(\phi_{0})\right]\\
b & =-\frac{1}{(2\pi)^{2d}}\frac{f'''(\phi_{0})}{2},
\end{align}
and $b>0$ by assumption. Note that $\psi(\mathbf{k})=\psi^{\ast}(-\mathbf{k})$,
because $\psi(\mathbf{r})$ is real-valued. Also, $A(|\mathbf{k}|)$
is real because $A(|\mathbf{x}|)$ is real and symmetric.

The cost of small perturbations about the liquid
phase, $\psi(\mathbf{k})=\phi-\phi_{0}$, scales as $A(|\mathbf{k}|)^{1/2}|\psi(\mathbf{k})|$.
The largest fluctuations occur at the angular frequency $|\mathbf{k}| = k_{0}$ that minimizes
$A(|\mathbf{k}|)$. Our interest is nucleation into a modulated phase,
and we choose the potential $\Lambda(\mathbf{x})$ such that $k_{0}>0$.
We expand about this minimum, 
\begin{equation}
A(|\mathbf{k}|)=\epsilon+\sigma^{2}(|\mathbf{k}|-k_{0})^{2}+\mathcal{O}((|\mathbf{k}|-k_{0})^{3})\label{eq:a},
\end{equation}
where 
\begin{align}
\epsilon & =  A(k_{0})\label{eq:epsilon}\\
\sigma^{2} & =  A''(k_{0})/2.\label{eq:sigma}
\end{align}
The parameter $\epsilon$ determines the stability
of the system. When $\epsilon<0$ the liquid phase is unstable to
fluctuations $\psi(\mathbf{k})$ with angular frequency $|\mathbf{k}|=k_{0}$. For $\epsilon>0$
sufficiently small, the liquid phase is metastable. We study nucleating
droplets near the limit of metastability (spinodal) of the liquid
phase, where 
$
0<\epsilon\ll1$.
In this limit, we will show that the truncations in Eqs.~\eqref{eq:free_psi}--\eqref{eq:a}
are self-consistent. That is, near the liquid phase spinodal, only
the small set of parameters $\{\epsilon,\sigma,k_{0},b\}$ is relevant.
We assume that nucleation is controlled by the saddle
point approximation~\cite{Cahn59,Langer67}, which is valid when
$R^{d}\gg1$ such that the metastable phase is long-lived~\cite{Klein86}.

We seek nucleating droplets as saddle point solutions of the Euler-Lagrange
equation expressed in Fourier space,
\begin{equation}
A(|\mathbf{k}|)\psi(\mathbf{k})=b(\psi\ast\psi)(\mathbf{k})+O(\psi^{3}).\label{eq:EL_psi}
\end{equation}
Nucleation occurs through localized droplet objects within a background
of the liquid phase ($\psi=0$). In this work, we compare two types
of nucleating droplets. The first type (introduced by Klein and Leyvraz~\cite{Klein86},
building on the symmetry arguments of Alexander and McTague~\cite{Alexander78}),
is a ramified droplet with a lattice structure at its core. The second
type (not previously considered) is a spherically symmetric solution
with concentric shells of modulated density. We will call these two
types lattice structure and onion structure droplets.

Our strategy is as follows. We propose a scaling ansatz for both lattice structure
and onion structure droplet types. By substituting these ansatzes into the Euler-Lagrange
equation, we show that both may be self-consistent solutions.
However, we find that the free energies of the two droplet types scale
differently. Our analysis, which is valid near the spinodal, predicts 
that lattice structure droplets are favored in two dimensions, while onion structure droplets
are favored in three and higher dimensions.

\section{Lattice structure droplets\label{sec:ls_droplets}}

\begin{figure*}
\includegraphics[scale=0.5]{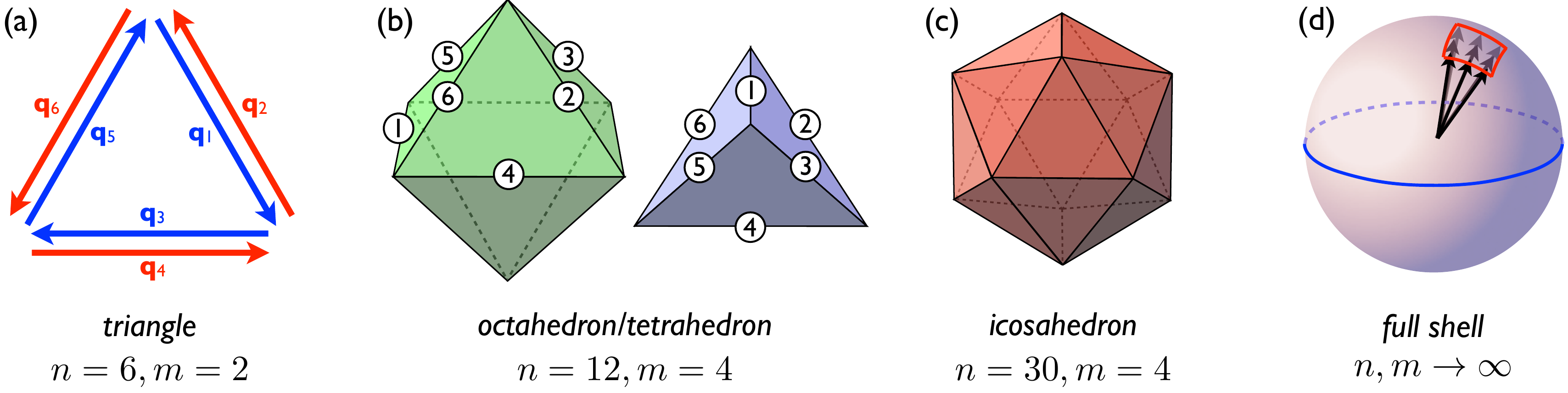}\caption{\label{fig:symmetries}Possible symmetric sets of reciprocal lattice
vectors satisfying the Alexander-McTague condition of equilateral
triangles. (a) The single triangle has $n=6$ edge vectors. For each
$\mathbf{q}_{i}$, there are $m=2$ ordered pairs $(\mathbf{q}_{j},\mathbf{q_{k}})$
such that $\mathbf{q}_{j}+\mathbf{q}_{k}=\mathbf{q}_{i}$. (b) The
octahedron and tetrahedron have equivalent sets of $n=12$ edge vectors.
Each edge lies on two faces, so $m=4$. The edge vectors correspond
to a bcc crystal in direct space. (c) The icosahedron is the only
remaining polyhedron composed of equilateral triangles. The edge vectors
correspond to a quasicrystal in direct space. (d) The complete set
of equal magnitude vectors corresponds to an onion structure in direct
space, and describes nucleation near the spinodal in $d\geq3$ dimensions.}
\end{figure*}
Following Klein and Leyvraz~\cite{Klein86}, we seek lattice structure droplet solutions to the Euler-Lagrange equation. We work with
the ansatz 
\begin{equation}
\psi_{\mathrm{LS}}(\mathbf{k})=\frac{\epsilon}{bm}\sum_{i=1}^{n}\left(\frac{\sigma}{\epsilon^{1/2}}\right)^{d}f_{i}\left(\frac{\sigma}{\epsilon^{1/2}}(\mathbf{k}-\mathbf{q}_{i})\right),\label{eq:psi_ls}
\end{equation}
valid at leading order in $\epsilon\ll1$, the distance
from the spinodal. The parameters $k_{0}$, $\sigma$, and $b$ are
as before, while the constant $m$ and functions $f_{i}(\mathbf{k})$
are to be determined. The $n$ reciprocal lattice vectors $\{\mathbf{q}_{i}\}$
all have magnitude $|\mathbf{q}_{i}|=k_{0}$, consistent with minimal
droplet free energy cost, Eq.~\eqref{eq:f2}.

The functions $f_{i}$ are peaked at the origin and scaled such that,
in the spinodal limit, 
\begin{equation}
\psi_{\mathrm{LS}}(\mathbf{k})\rightarrow\frac{\epsilon}{bm}\sum_{i=1}^{n}\delta(\mathbf{k}-\mathbf{q}_{i})\quad\mathrm{as}\quad\epsilon\rightarrow0,\label{eq:ls_scaling}
\end{equation}
where $\delta(\mathbf{k})$ is the Dirac $\delta$-function. For $\epsilon$
positive but small, the ansatz $\psi_{\mathrm{LS}}(\mathbf{k})$ is
a sum of highly peaked terms.

Following Alexander and McTague~\cite{Alexander78}, we use symmetry
arguments to constrain the possible symmetries of the reciprocal vectors
$\{\mathbf{q}_{i}\}$. For this argument, we work with the asymptotic
representation of Eq.~\eqref{eq:ls_scaling}, valid at
leading order in $\epsilon$~\cite{Klein01}. We insert the ansatz
$\psi_{\mathrm{LS}}(\mathbf{k})$ into the Euler-Lagrange equation~\eqref{eq:EL_psi},
and expand the right-hand side using the identity 
\begin{equation}
g(\mathbf{k}-\mathbf{q}_{i})\ast h(\mathbf{k}-\mathbf{q}_{j})=(g\ast h)(\mathbf{k}-\mathbf{q}_{i}-\mathbf{q}_{j}).\label{eq:conv}
\end{equation}

The left-hand side of Eq.~\eqref{eq:EL_psi} is approximately $A(k_{0})\psi_{\mathrm{LS}}(\mathbf{k})=\epsilon\psi_{\mathrm{LS}}(\mathbf{k})$,
a sum of $\delta$-functions at the reciprocal vectors $\{\mathbf{q}_{i}\}$.
The right-hand side contains approximate $\delta$-functions at $\{\mathbf{q}_{j}+\mathbf{q}_{k}\}$
for all pairs $j,k$. The $i$th left-hand side term must be matched
by corresponding ones on the right-hand side,
\begin{equation}
\epsilon\left[\frac{\epsilon}{bm}\delta(\mathbf{k}-\mathbf{q}_{i})\right]=b\!\sum_{\substack{j,k=1\\
\mathbf{q}_{j}+\mathbf{q}_{k}=\mathbf{q}_{i}
}
}^{n}\left(\frac{\epsilon}{bm}\right)^{2}(\delta\ast\delta)(\mathbf{k}-\mathbf{q}_{j}-\mathbf{q}_{k}).\label{eq:EL_ls_delta}
\end{equation}
Using $(\delta*\delta)(\mathbf{k})=\delta(\mathbf{k})$, we find that
every $\mathbf{q}_{i}$ on the left-hand side must be matched by $m$ ordered
pairs $(\mathbf{q}_{j},\mathbf{q}_{k})$ satisfying $\mathbf{q}_{j}+\mathbf{q}_{k}=\mathbf{q}_{i}$
on the right-hand side. What about unmatched terms on the right-hand side of Eq.~\eqref{eq:EL_psi},
for which $\mathbf{p}=\mathbf{q}_{j}+\mathbf{q}_{k}$ is different
than all $\mathbf{q}_{i}$? There are two possibilities. If $|\mathbf{p}|\neq k_{0}$,
then these terms can be accounted for by higher order corrections,
$\psi_{\mathrm{LS}}\rightarrow\psi_{\mathrm{LS}}+\delta\psi_{\mathrm{LS}}$.
Namely, we balance the left-  and right-hand sides, $A(|\mathbf{p}|)\delta\psi(\mathbf{p})\sim(\psi*\psi)(\mathbf{p})$, 
where $\psi(\mathbf{p})\sim\epsilon$ and $A(|\mathbf{p}|)\sim1$,
and find that $\delta\psi(\mathbf{p})\sim\epsilon^{2}$ is a higher
order correction. The second case, $|\mathbf{p}|=k_{0}$, is disallowed
because $A|\mathbf{p}|\sim\epsilon$, and it is \emph{not} possible
to satisfy the Euler-Lagrange equation.

We conclude that the reciprocal lattice vectors $\{\mathbf{q}_{i}\}$,
each of magnitude $k_{0}$, must form equilateral triangles. The possible
closed, symmetric sets of equilateral triangles are shown in Fig.~\ref{fig:symmetries}.
The simplest solution is the set of $n=6$ edge vectors of a single triangle.
In direct space, this solution represents a triangular lattice (in 2d)
or close packed rod-like structures (in 3d). There are only three polyhedra
composed of equilateral triangles. The $n=12$ unique edge vectors
of the octahedron and tetrahedron are equivalent, and correspond to
a bcc lattice in direct space. The $n=30$ unique edge vectors of
the icosahedron correspond to a quasicrystal in direct space. These
finite sets complete the classification given by Alexander and McTague~\cite{Alexander78},
and are the basis of the lattice structure droplets we will describe in the remainder of this section.
In Sec.~\ref{sec:os_droplets} we consider instead the \emph{continuum}
set of vectors $\{\mathbf{q}\}$ with magnitude $|\mathbf{q}|\approx k_{0}$.

For simplicity, we used the asymptotic scaling of Eq.~\eqref{eq:ls_scaling}
to derive the Alexander-McTague symmetry constraints. Now we return
to the full ansatz Eq.~\eqref{eq:psi_ls}, where $\psi_{\mathrm{LS}}(\mathbf{k})$
has sharp but non-singular peaks at $\{\mathbf{q}_{i}\}$. By symmetry,
we may focus on one particular $\mathbf{q}_{i}$ without loss of generality.
For $\mathbf{k}$ in the neighborhood of $\mathbf{q}_{i}$, we approximate
\begin{equation}
|\mathbf{k}|-k_{0}=\mathbf{k}\cdot\left(\frac{\mathbf{q}_{i}-\mathbf{k}}{k_{0}}\right)+\mathcal{O}(\delta k).
\end{equation}
We will find, self consistently, that the relevant scale is
\begin{equation}
\delta k=|\mathbf{k}-\mathbf{q}_{i}|\sim\epsilon^{1/2}/\sigma.\label{eq:delta_k}
\end{equation}
In particular, we may construct a vector of order unity,
\begin{equation}
\mathbf{v}=\frac{\sigma}{\epsilon^{1/2}}(\mathbf{k}-\mathbf{q}_{i}).\label{eq:v}
\end{equation}
We insert these definitions into Eq.~\eqref{eq:a} and find
\begin{equation}
A(|\mathbf{k}|)=\epsilon(1+(\mathbf{v}\cdot\mathbf{q}_{i}/k_{0})^{2})+\mathcal{O}(\epsilon^{3/2}),\label{eq:a_qi}
\end{equation}
when $\delta k \sim \epsilon^{1/2}$.

At leading order in $\epsilon$, the Euler-Lagrange equation~\eqref{eq:EL_psi}
for the full ansatz, Eq.~\eqref{eq:psi_ls}, becomes,
\begin{equation}
(1+(\mathbf{v}\cdot\mathbf{q}_{i}/k_{0})^{2})f_{i}(\mathbf{v})=\frac{1}{m}\!\sum_{\substack{j,k=1\\
\mathbf{q}_{j}+\mathbf{q}_{k}=\mathbf{q}_{i}
}
}^{n}(f_{j}\ast f_{k})(\mathbf{v}),\label{eq:EL_ls_f}
\end{equation}
where $(f_{j}\ast f_{k})$ denotes convolution with respect to $\mathbf{v}$,
namely $\int f_{j}(\mathbf{v}')f_{k}(\mathbf{v}-\mathbf{v}')\mathd^{d}\mathbf{v}'$.
Solutions to Eq.~\eqref{eq:EL_ls_f}, if they exist, would self-consistently
justify the ansatz $\psi_{\mathrm{LS}}(\mathbf{k})$ as well as the
restriction to small $\delta k$.

We ignore the trivial solution $f_{i}(\mathbf{v})=0$, which represents no deviation from
the metastable liquid phase  $\psi(\mathbf{k})=0$.

Another trivial solution, $f_{i}(\mathbf{v})=\delta(\mathbf{v})$,
corresponds to a direct space lattice of triangular, bcc, or quasi-crystal
symmetry [\emph{cf.}\ Figs.~\ref{fig:symmetries}(a)--(c)]. These lattice
solutions are known to be unstable, and therefore do not represent
an equilibrium phase~\cite{Klein01}. Furthermore, their extensive
free energy cost excludes their interpretation as nucleating droplets.

We seek smooth, non-singular solutions $f_{i}(\mathbf{v})$ of Eq.~\eqref{eq:EL_ls_f}.
For such solutions, the ansatz $\psi_{\mathrm{LS}}(\mathbf{k})$ would
potentially represent a nucleating droplet appearing in the metastable
liquid phase. In direct space, this droplet would have spatial scale
$\ell\sim R\sigma/\epsilon^{1/2}$. We expect droplet solutions to
inherit the symmetries of the reciprocal lattice vectors $\{\mathbf{q}_{i}\}$.
In particular, all point group symmetries $\mathcal{R}\{\mathbf{q}_{i}\}=\{\mathbf{q}_{i}\}$
should be represented by the droplet, $\psi_{\mathrm{LS}}(\mathbf{k})=\psi_{\mathrm{LS}}(\mathcal{R}\mathbf{k})$,
and the envelope functions $\{f_{i}\}$ should satisfy,
\begin{equation}
f_{i}(\mathbf{k})=f_{j}(\mathcal{R}\mathbf{k})\quad\mathrm{if}\quad\mathbf{q}_{j}=\mathcal{R}\mathbf{q}_{i}.\label{eq:rot}
\end{equation}

In direct space, the full nucleating droplet solution will exhibit
anisotropic faceting due to the term $\mathbf{v}\cdot\mathbf{q}_{i}$
appearing on the left-hand side of Eq.~\eqref{eq:EL_ls_f}. Analytical solution
appears difficult. The substitution $(\mathbf{v}\cdot\mathbf{q}_{i}/k_{0})^{2}\rightarrow\mathbf{v}^{2}$
would restore isotropy and yield a theory analogous to that of ferromagnetic
nucleation, but approximation is uncontrolled~\cite{Unger84}.

\begin{figure}
\includegraphics[scale=0.48]{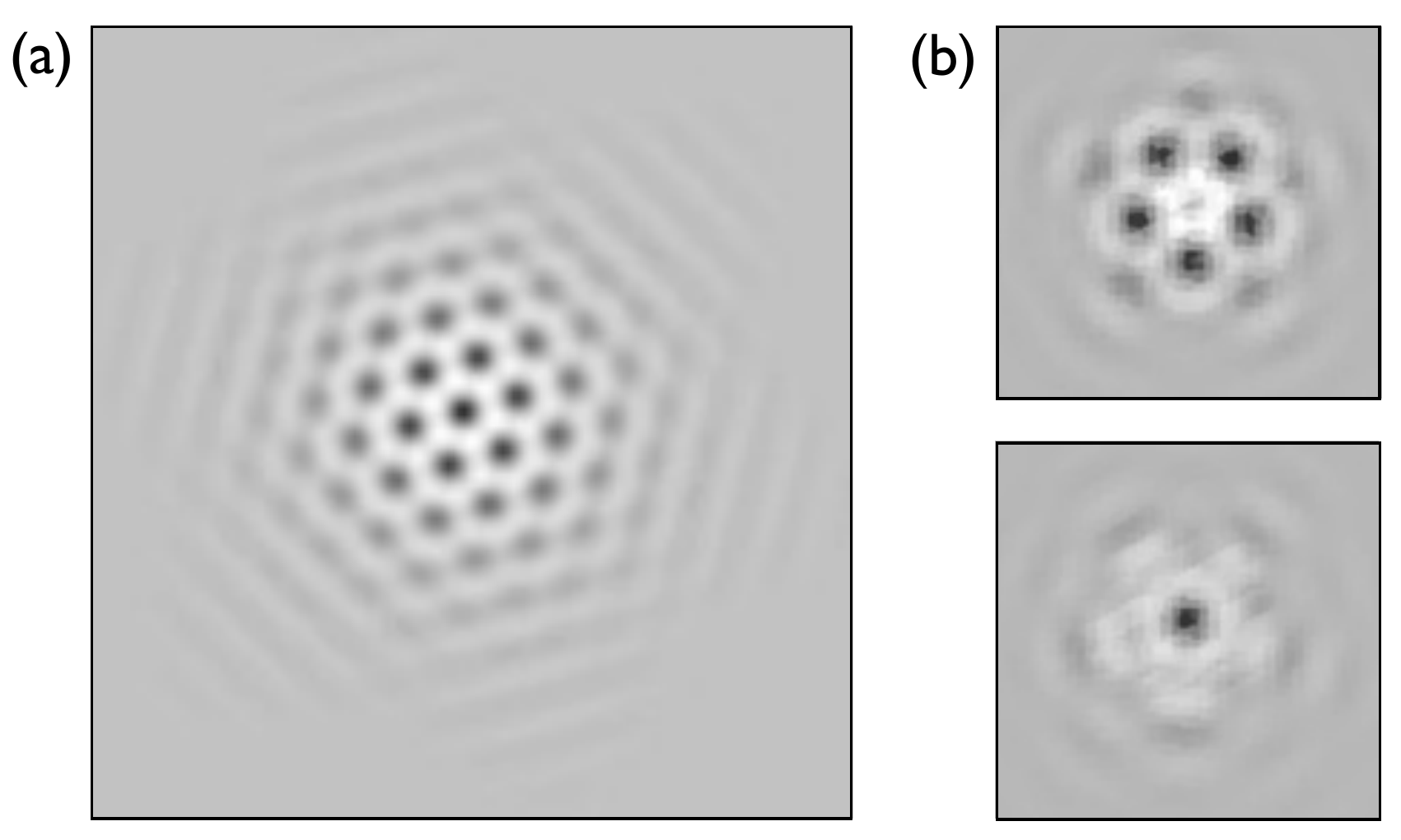}\caption{\label{fig:lattice}Free energy saddle points representing nucleating
droplets from the metastable liquid phase. (a) In two-dimensions,
nucleation always occurs through a droplet with triangular lattice
symmetry. (b) In three-dimensions, the symmetry of the droplet depends
on the distance $\epsilon$ from the spinodal. In numerical studies
of the clump model, Eqs.~\eqref{eq:step1} and~\eqref{eq:step2},  at intermediate $\epsilon$, we find a droplet
core containing 13 close packed spheres with icosahedral symmetry
(two slices are shown).}
\end{figure}

Numerical solutions to the non-truncated Euler-Lagrange equation~\eqref{eq:EL_phi}
are shown in Fig.~\ref{fig:lattice}. We applied the string method~\cite{E07,Backofen10,Ren13}
to find saddle points of minimum free energy for the clump model, Eqs.~\eqref{eq:step1} and \eqref{eq:step2}.
Up to a simple rescaling, the clump model is independent of the liquid
phase density $\phi_{0}$. The spinodal temperatures are $T_{s}^{\rm 2d}=0.13228\, \phi_{0}$
and $T_{s}^{\rm 3d}=0.08617\, \phi_{0}$ in two and three dimensions, respectively.
We used $\epsilon=(T-T_{s})/\phi_{0}=0.01$ to calculate the droplets
shown in Fig.~\ref{fig:lattice}. In two dimensions we find, for
all $\epsilon$, that the nucleating droplets have an extended core
with triangular lattice symmetry. For small $\epsilon$, we observe
scaling consistent with the ansatz $\psi_{\mathrm{LS}}(\mathbf{k})$.
In three dimensions and intermediate values of $\epsilon$, we observe
droplets with an icosahedral core. This result is
non-universal in the sense that other models may have a different
structure. Closer to the spinodal ($\epsilon\lesssim0.005$ for the
clump model) we instead find the onion structure droplets that we describe in the next section.

\section{Onion structure droplets\label{sec:os_droplets}}

In this section, we propose a new kind of nucleating droplet composed
of \emph{all} vectors $\{\mathbf{q}\}$ satisfying $|\mathbf{q}|=k_{0}$.
This continuum set may be interpreted as a degenerate solution to
the Alexander-McTague symmetry constraints, Fig.~\ref{fig:symmetries}(d).
In direct space, this solution becomes a spherically symmetric, radially
modulated density field that resembles the layers of an onion.

Our full ansatz for onion structure droplets is, 
\begin{equation}
\psi_{\mathrm{OS}}(|\mathbf{k}|)=\gamma g\left(\frac{\sigma}{\epsilon^{1/2}}(|\mathbf{k}|-k_{0})\right).\label{eq:psi_os}
\end{equation}
The parameters $k_{0}$ and $\sigma$ are as before. At leading order
in $0<\epsilon\ll1$, we will self-consistently solve for $\gamma$
and $g(u)$. The latter is centered and normalized such that 
\begin{equation}
\int\! g(u)\mathd u=1,\label{eq:g_norm}
\end{equation}
and 
\begin{equation}
\psi_{\mathrm{OS}}(|\mathbf{k}|)\rightarrow\frac{\epsilon^{1/2}\gamma}{\sigma}\delta(|\mathbf{k}|-k_{0})\quad\mathrm{as}\quad\epsilon\rightarrow0.\label{eq:os_scaling}
\end{equation}
Whereas $\psi_{\mathrm{LS}}(\mathbf{k})$ was composed of sharp peaks
at the discrete set of vectors $\{\mathbf{q}_{i}\}$ [\emph{cf.}\ Eq.~\eqref{eq:ls_scaling}],
the new solution $\psi_{\mathrm{OS}}(|\mathbf{k}|)$ shows a non-zero
spectral density along the full shell of radius $k_{0}$.

\begin{figure}

\includegraphics[scale=0.45]{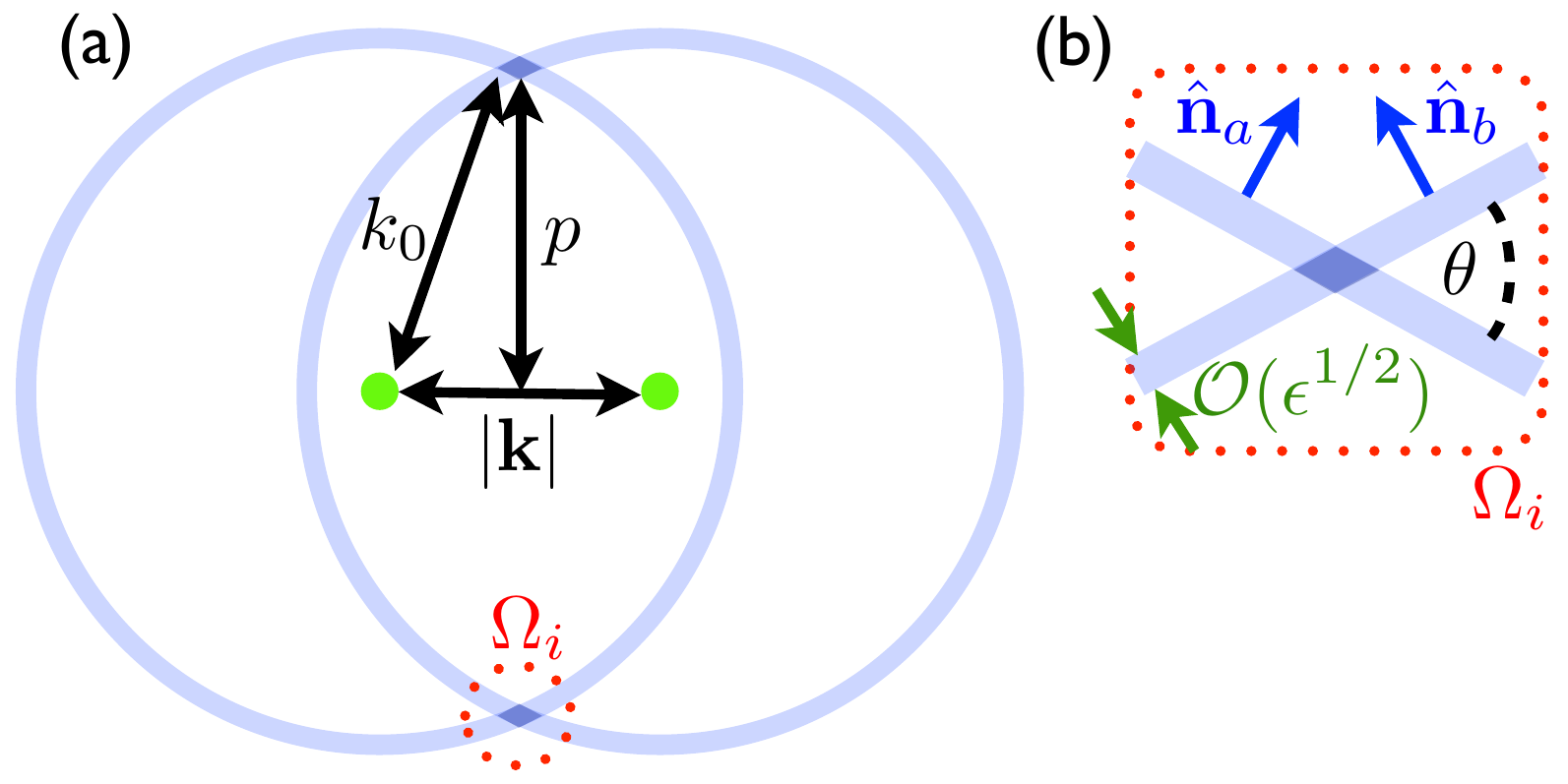}\caption{\label{fig:conv}(a) The convolution $(\psi_{\mathrm{OS}}\ast\psi_{\mathrm{OS}})(|\mathbf{k}|)$
represents an integral over the product of two thin spherical shells.
The overlap area is a lower dimensional sphere of radius $p$. (b)
When the shell thickness $\mathcal{O}(\epsilon^{1/2})$ is small compared
to $p$, the shell curvature can be ignored in each sub-region $\Omega_{i}$
of overlap. The angle between surface normals $\hat{\mathbf{n}}_{a}$
and $\hat{\mathbf{n}}_{b}$ is given by $\sin\theta=kp/k_{0}^{2}$.}
\end{figure}

We anticipate restricting $|\mathbf{k}|$ to the neighborhood of $k_{0}$,
such that
\begin{equation}
u=\frac{\sigma}{\epsilon^{1/2}}(|\mathbf{k}|-k_{0})\label{eq:u}
\end{equation}
 is of order unity [\emph{cf.}\ Eq.~\eqref{eq:v}]. In particular, for
small 
$
\delta k=|\mathbf{k}|-k_{0}$
we may write
\begin{equation}
A(|\mathbf{k}|)=\epsilon(1+u^{2})+\mathcal{O}(\epsilon^{3/2}).
\end{equation}
The left-hand side of Eq.~\eqref{eq:EL_psi} becomes
\begin{equation}
A(|\mathbf{k}|)\psi_{\mathrm{OS}}=\epsilon(1+u^{2})\gamma g(u)+\mathcal{O}(\epsilon^{3/2}).\label{eq:EL_OS_l}
\end{equation}
We now turn to the right-hand side of Eq.~\eqref{eq:EL_psi}. As a first approximation,
we use the asymptotic representation of Eq.~\eqref{eq:os_scaling},
\begin{equation}
\begin{split}(\psi_{\mathrm{OS}} & \ast\psi_{\mathrm{OS}})(|\mathbf{k}|)\\
 & \approx\frac{\epsilon\gamma^{2}}{\sigma^{2}}\!\int\delta(|\mathbf{k}'|-k_{0})\delta(|\mathbf{k}-\mathbf{k}'|-k_{0})\mathd^{d}\mathbf{k}'.
\end{split}
\label{eq:conv_scaling}
\end{equation}
This convolution corresponds to the integrated overlap between the
two shells, illustrated in Fig.~\ref{fig:conv}(a). The overlap area
is $p^{d-2}S_{d-2}$ where 
\begin{equation}
p=\sqrt{k_{0}^{2}-|\mathbf{k}|^{2}/4},
\end{equation}
 and $S_{n}$ is the surface area of the $n$-sphere, 
\begin{equation}
S_{n}=\frac{2\pi^{(n+1)/2}}{\Gamma[(n+1)/2]}=
\begin{cases}
2 &   n=0\\
2\pi & n=1\\
4\pi &  n=2
\end{cases}.\label{eq:nsphere}
\end{equation}
To evaluate $\psi_{\mathrm{OS}}\ast\psi_{\mathrm{OS}}$ more precisely,
we return to the full ansatz, Eq.~\eqref{eq:psi_os}. In a local region
$\Omega_{i}$ of overlap, Fig.~\ref{fig:conv}(b), we treat the shells
as (hyper)planes with surface normals satisfying $\hat{\mathbf{n}}_{a}\cdot\hat{\mathbf{n}}_{b}=\cos\theta$.
The integral in Eq.~\eqref{eq:conv_scaling} becomes a sum over many
such regions,
\begin{equation}
(\psi_{\mathrm{OS}}\ast\psi_{\mathrm{OS}})(|\mathbf{k}|)=\gamma^{2}\sum_{i}\int_{\Omega_{i}}g\left(v_{a}\right)g\left(v_{b}\right)\mathd^{d}\mathbf{k}'+\mathcal{O}(\epsilon^{3/2}),\label{eq:conv_general}
\end{equation}
where

\begin{equation}
v_{\{a,b\}}=\frac{\sigma}{\epsilon^{1/2}}\mathbf{k}'\cdot\hat{\mathbf{n}}_{\{a,b\}}.
\end{equation}
Curvature effects are negligible because the shells have width $\mathcal{O}(\epsilon^{1/2})$.
In each region $\Omega_{i}$ we select an orthogonal coordinate system
$\{\hat{\mathbf{e}}_{1},\hat{\mathbf{e}}_{2}, \ldots,\hat{\mathbf{e}}_{d}\}$
in which the surface normals are $\hat{\mathbf{n}}_{a}=\hat{\mathbf{e}}_{1}$
and $\hat{\mathbf{n}}_{b}=\cos\theta\,\hat{\mathbf{e}}_{1}+\sin\theta\,\hat{\mathbf{e}}_{2}$.
The integrand then depends only on the first two components, $k'_{1}$
and $k'_{2}$, of $\mathbf{k}'$. After two applications of the normalization
condition, Eq.~\eqref{eq:g_norm}, we find
\begin{equation}
\begin{split}\int g\left(v_{a}\right)g\left(v_{b}\right)\mathd k_{1}'\mathd k_{2}' & =\frac{\epsilon^{1/2}}{\sigma\sin\theta}\!\int g\left(v_{a}\right)\mathd k_{1}'\\
 & =\frac{\epsilon}{\sigma^{2}\sin\theta}.
\end{split}
\end{equation}
The remaining ($d-2$) orthogonal integrations, summed over all regions $\Omega_i$, evaluate to the area
of overlap,
\begin{equation}
\sum_{i}\!\int_{\Omega_{i}}\prod_{j=2}^{d}\mathd k_{j}'=p^{d-2}S_{d-2}.
\end{equation}
The full convolution becomes
\begin{equation}
(\psi_{\mathrm{OS}}\ast\psi_{\mathrm{OS}})(|\mathbf{k}|)=\frac{\epsilon\gamma^{2}}{\sigma^{2}}\frac{p^{d-2}}{\sin\theta}S_{d-2}+\mathcal{O}(\epsilon^{3/2}).
\end{equation}
Inspection of Fig.~\ref{fig:conv} gives the geometric identity,
\[
\sin\theta=kp/k_{0}^{2},
\]
and, at leading order in $\delta k\sim\epsilon^{1/2}$,
\begin{equation}
\begin{split}(\psi_{\mathrm{OS}}\ast\psi_{\mathrm{OS}})(|\mathbf{k}|) & =\frac{\epsilon\gamma^{2}}{\sigma^{2}}\frac{p^{d-3}k_{0}^{2}}{k}S_{d-2}+\mathcal{O}(\epsilon^{3/2})\\
 & =\frac{\epsilon\gamma^{2}}{\sigma^{2}}\left(\frac{3}{4}\right)^{\frac{d-3}{2}}k_{0}^{d-2}S_{d-2}+\mathcal{O}(\epsilon^{3/2}).
\end{split}
\label{eq:EL_OS_r}
\end{equation}
With Eqs.~\eqref{eq:EL_OS_l} and \eqref{eq:EL_OS_r}, we satisfy the
Euler-Lagrange equation~\eqref{eq:EL_psi} at leading order in $\epsilon$
with
\begin{align}
g(u) & =\frac{1}{\pi(1+u^{2})}\label{eq:g_sol}\\
\gamma & =\frac{\sigma^{2}}{b\pi\left(3/4\right)^{\frac{d-3}{2}}k_{0}^{d-2}S_{d-2}}.\label{eq:gamma_sol}
\end{align}
Note that $g$ is properly normalized, Eq.~\eqref{eq:g_norm}. This
self-consistent solution justifies the ansatz $\psi_{\mathrm{OS}}(|\mathbf{k}|)$
at small $\delta k$. Our final expression for the onion structure
droplets in Fourier space is, 
\begin{equation}
\psi_{\mathrm{OS}}(|\mathbf{k}|)=\frac{\sigma^{2}}{b\pi^{2}\left(3/4\right)^{\frac{d-3}{2}}k_{0}^{d-2}S_{d-2}}(1+u^{2})^{-1},\label{eq:os_sol_k}
\end{equation}
where $u$ is defined in Eq.~\eqref{eq:u}.

\begin{figure}
\includegraphics[scale=0.35]{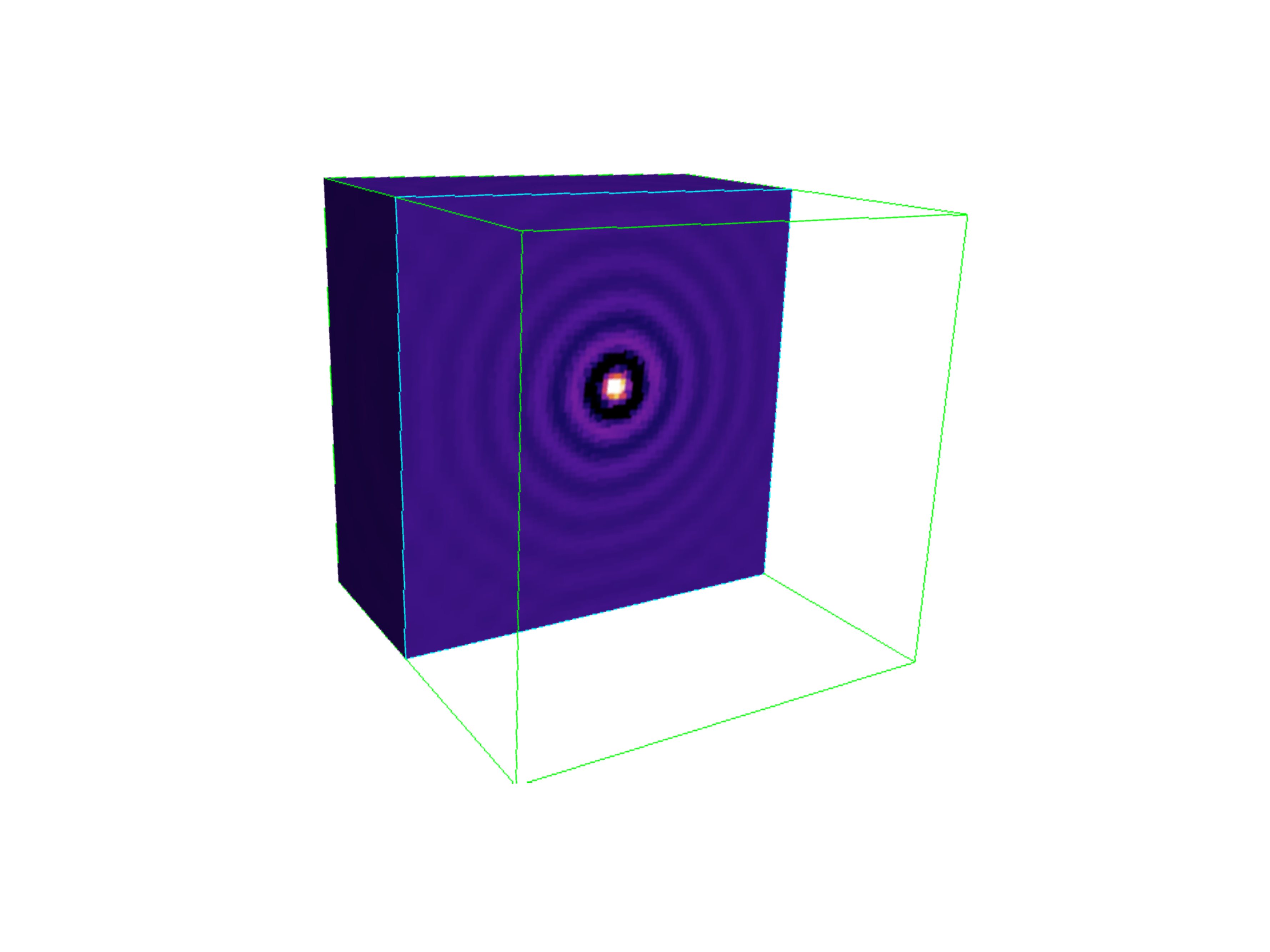}\caption{\label{fig:onion}In three dimensions and higher, nucleation near
the liquid spinodal ($\epsilon\ll1$) occurs via the spherically symmetric
onion structure droplet. Among all saddle points, this droplet
has minimum free energy.}
\end{figure}

We can also express the onion structure droplets in real space by the inverse
Fourier transformation,
\begin{equation}
\psi_{\mathrm{OS}}(|\mathbf{r}|)=(2\pi)^{-d}\!\int e^{-i\mathbf{k}\cdot\mathbf{r}}\psi_{\mathrm{OS}}(|\mathbf{k}|)\mathd^{d}\mathbf{k}.\label{eq:os_real1}
\end{equation}
In the following, we use the shorthand notation $r=|\mathbf{r}|$.
In spherical coordinates we have
\begin{equation}
\mathd^{d}\mathbf{k}=\left[\mathd k\right] \left[(k\sin\theta)^{d-2}\mathd\theta\right] \left[k\mathd^{d-2}\phi\right],
\end{equation}
where $k$ is the distance from the origin and $\theta$ is the zenith
angle. The remaining $d-2$ surface angles $\phi$ satisfy $\int\mathd^{d-2}\phi=S_{d-2}$,
leaving only two integrals,
\begin{align}
\psi_{\mathrm{OS}}(r) & =\alpha\!\int_{0}^{\pi}I(\theta)(\sin\theta)^{d-2}\mathd\theta\label{eq:os_real2}\\
I(\theta) & =\!\int_{0}^{\infty}k^{d-1}e^{-ikr\cos\theta}\psi_{\mathrm{OS}}(k)\mathd k,
\end{align}
where
\begin{equation}
\alpha=S_{d-2}(2\pi)^{-d}.\label{eq:alpha_def}
\end{equation}
Because $\psi_{\mathrm{OS}}(k)$ is sharply peaked at $k=k_{0}$,
we may extend the integral bounds to the entire real line. We insert
the ansatz, Eq.~\eqref{eq:psi_os}, keep only leading order terms
in $\epsilon$ (at all $r$), and obtain
\begin{equation}
I(\theta)=\beta\!\int_{-\infty}^{+\infty}\exp\left[-i\left(k_{0}+\frac{\epsilon^{1/2}}{\sigma}u\right)r\cos\theta\right]g(u)\mathd u,
\end{equation}
where
\begin{equation}
\beta=k_{0}^{d-1}\frac{\epsilon^{1/2}\gamma}{\sigma},\label{eq:beta_def}
\end{equation}
and $g(u)$ is given in Eq.~\eqref{eq:g_sol}. Using the identity
\begin{equation}
\int_{-\infty}^{+\infty}du\frac{e^{icu}}{\pi(1+u^{2})}=e^{-|c|},
\end{equation}
we obtain
\begin{equation}
I(\theta)=\beta\exp\left(-ik_{0}r\cos\theta-\frac{\epsilon^{1/2}}{\sigma}r|\cos\theta|\right).
\end{equation}
Equation~\eqref{eq:os_real2} becomes
\begin{equation}
\begin{split}\psi & _{\mathrm{OS}}(r)=\\
 & \alpha\beta \!\int_{0}^{\pi}\exp\left(-ik_{0}r\cos\theta-\frac{\epsilon^{1/2}}{\sigma}r|\cos\theta|\right)(\sin\theta)^{d-2}\mathd\theta.
\end{split}
\label{eq:os_real3}
\end{equation}
We may use the integral representation of the Bessel function of the first
kind, 
\begin{equation}
J_{\nu}(r)=\frac{\left(r/2\right)^{\nu}}{\pi^{1/2}\Gamma(\nu+1/2)}\int_{0}^{\pi}e^{-ir\cos\theta}\sin^{2\nu}\theta d\theta,
\end{equation}
to obtain a result valid when $r\ll\epsilon^{-1/2}$, 
\begin{equation}
\psi_{\mathrm{OS}}(r)\approx\alpha\beta\frac{\pi^{1/2}\Gamma(\frac{d-1}{2})}{\left(k_{0}r/2\right)^{d/2-1}}J_{d/2-1}(k_{0}r).\label{eq:os_real_sol1}
\end{equation}
The neglected $\epsilon$ dependence in Eq.~\eqref{eq:os_real3}
damps $\psi_{\mathrm{OS}}(r)$ on the scale $r\sim\epsilon^{-1/2}$.
In $d=3$  the full leading order result is 
\begin{equation}
\psi_{\mathrm{OS}}^{d=3}=\exp[-\frac{\epsilon^{1/2}}{\sigma}r]\left(\alpha\beta\frac{2\sin k_{0}r}{k_{0}r}\right),\label{eq:os_real_sol2}
\end{equation}
which is equivalent to Eq.~\eqref{eq:os_real_sol1} with an exponential
damping factor. By substituting definitions from Eqs.~\eqref{eq:gamma_sol}, \eqref{eq:alpha_def},
and \eqref{eq:beta_def}, we obtain our final droplet solution, 
\begin{equation}
\psi_{\mathrm{OS}}^{d=3}(r)=\exp[-\epsilon^{1/2}\sigma^{-1}r]\frac{\epsilon^{1/2}\sigma k_{0}}{4\pi^{4}b}\left(\frac{\sin k_{0}r}{k_{0}r}\right).\label{eq:os_real_sol3}
\end{equation}
We numerically calculated the free energy saddle point using non-truncated
Euler-Lagrange equation~\eqref{eq:EL_phi} for the clump model
near the spinodal, $\epsilon\lesssim0.005$, and found the onion structure
droplet $\psi_{\mathrm{OS}}(r)$ shown in Fig.~\ref{fig:onion}.

\section{Droplet free energies\label{sec:energies}}

When large fluctuations about the metastable liquid phase are rare,
nucleation occurs via the saddle point of minimum free energy. Both
lattice structure [$\psi_{\mathrm{LS}}$, Eq.~\eqref{eq:psi_ls}] and
onion structure [$\psi_{\mathrm{OS}}$, Eq.~\eqref{eq:psi_os}] solutions
are possible nucleating droplets. In this section, we compute their
free energies to predict the preferred nucleating droplet structure.

The $\epsilon$-scaling of $\psi_{\mathrm{LS}}$ and $\psi_{\mathrm{OS}}$
ensures that $\mathcal{O}(\psi^{4})$ free energy terms are negligible
in the limit $\epsilon\ll1$. The two terms $F_{2}$ and $F_{3}$
in Eq.~\eqref{eq:free_psi} remain. For droplets $\psi$ satisfying
the Euler-Lagrange equation \eqref{eq:EL_psi}, the two terms are related,
\begin{equation}
\begin{split}F_{3} & =-\frac{b}{3}R^{d}\!\int\psi(\mathbf{k})\psi(\mathbf{k}')\psi(-\mathbf{k}-\mathbf{k}')\mathd^{d}\mathbf{k}\mathd^{d}\mathbf{k}'\\
 & =-\frac{1}{3}R^{d}\!\int\psi(\mathbf{k})[b(\psi\ast\psi)(-\mathbf{k})]\mathd^{d}\mathbf{k}\\
 & =-\frac{2}{3} \frac{1}{2}R^{d}\!\int\psi(\mathbf{k})[A(|\mathbf{k}|)\psi(-\mathbf{k})]\mathd^{d}\mathbf{k} =-\frac{2}{3}F_{2}.
\end{split}
\end{equation}
The total free energy becomes
\begin{equation}
F=F_{2}+F_{3}=\frac{1}{3}F_{2}.\label{free_reduced}
\end{equation}

We now evaluate the free energy of onion structure droplets,
\begin{equation}
F[\psi_{\mathrm{OS}}]=\frac{R^{d}}{6}\!\int A(|\mathbf{k}|)\psi_{\mathrm{OS}}(|\mathbf{k}|)^{2}\mathd^{d}\mathbf{k}.
\end{equation}
Because $\psi_{\mathrm{OS}}(|\mathbf{k}|)$ is sharply peaked at $k_{0}$,
we expand $|\mathbf{k}|=k_{0}+\delta k$ for $\delta k\sim\epsilon^{1/2}$,
truncate the expansion of $A|\mathbf{k}|$, Eq.~\eqref{eq:a}, and
extend the bounds of integration,
\begin{equation}
F[\psi_{\mathrm{OS}}]=\frac{R^{d}}{6}k_{0}^{d-1}S_{d-1}\!\int_{-\infty}^{+\infty}\epsilon(1+u^{2})\psi_{\mathrm{OS}}^{2}(k)\mathd k,
\end{equation}
where $u$ is defined in Eq.~\eqref{eq:u}. We insert the solution
$\psi_{\mathrm{OS}}(|\mathbf{k}|)$ from Eqs.~\eqref{eq:psi_os}, \eqref{eq:g_sol},
and \eqref{eq:gamma_sol} and evaluate the integral to get our final
result
\begin{equation}
F[\psi_{\mathrm{OS}}]=\frac{\epsilon^{3/2}\sigma^{3}R^{d}}{b^{2}k_{0}^{d-3}} \frac{S_{d-1}}{6\pi^{3}(3/4)^{d-3}S_{d-2}^{2}}.\label{eq:free_os}
\end{equation}

Next we consider  the lattice structure droplets. Assuming that solutions
to Eq.~\eqref{eq:EL_ls_f} exist, the droplet free energy is
\begin{equation}
F[\psi_{\mathrm{LS}}]=\frac{R^{d}}{6}\!\int A(|\mathbf{k}|) \psi_{\mathrm{LS}}(|\mathbf{k}|)^{2} \mathd^{d} \mathbf{k}.
\end{equation}
The ansatz $\psi_{\mathrm{LS}}(\mathbf{k})$ in Eq.~\eqref{eq:psi_ls}
is a sum of $n$ symmetrically-equivalent peaks $f_{i}$, Eq.~\eqref{eq:rot}.
At the $i$th peak, the relevant scale $\delta k=|\mathbf{k}-\mathbf{q}_{i}|$
is $\mathcal{O}(\epsilon^{1/2})$, and we can truncate the expansion
of $A(|\mathbf{k}|)$ in Eq.~\eqref{eq:a_qi}. The free energy from
all $n$ peaks becomes 
\begin{equation}
F[\psi_{\mathrm{LS}}]=R^{d}\frac{n}{6}\int\epsilon(1+(\mathbf{v}\cdot\mathbf{e}_{i})^{2})\left[\frac{\epsilon}{bm}\left(\frac{\sigma}{\epsilon^{1/2}}\right)^{d}f_{i}(\mathbf{v})\right]^{2}\mathd^{d}\mathbf{k},
\end{equation}
where $\mathbf{v}$ is defined in Eq.~\eqref{eq:v}. After a change
of variables and some algebra, we find
\begin{equation}
F[\psi_{\mathrm{LS}}]=\frac{\epsilon^{3-d/2}\sigma^{d}R^{d}}{b^{2}} \frac{cn}{6m^{2}},\label{eq:free_kl}
\end{equation}
with
\begin{equation}
c=\!\int(1+(\mathbf{v}\cdot\mathbf{q}_{i}/k_{0})^{2})f_{i}(\mathbf{v})^{2}\mathd^{d}\mathbf{v}.
\end{equation}
As before, $m$ represents the number of ordered pairs of reciprocal
lattice vectors $(\mathbf{q}_{j},\mathbf{q}_{k})$ that sum to a specific
$\mathbf{q}_{i}$. The possible values of $n$ and $m$ are listed
in Fig.~\ref{fig:lattice}. Although the solution $f_{i}$ of Eq.~\eqref{eq:EL_ls_f}
(if it exists) is not known analytically, by construction it is a
dimensionless function independent of the model parameters. It is
therefore natural to assume that $c$ is of order unity.

The free energy costs of the onion structure ($F[\psi_{\mathrm{OS}}]$
in Eq.~\eqref{eq:free_os}) and lattice structure ($F[\psi_{\mathrm{LS}}]$
in Eq.~\eqref{eq:free_kl}) droplets allow us to make predictions for
nucleation near the spinodal ($\epsilon\ll1$) when the metastable
liquid phase is long lived ($R\gg1$).
\begin{itemize}
\item In $d=1$ both $\psi_{\mathrm{OS}}\sim1/S_{d-2}$ and $\psi_{\mathrm{LS}}\sim1/m$
are ill-defined. Note that $m=0$ because a single wave vector cannot
form an equilateral triangle. This absence of nucleating droplets
is consistent with the continuous phase transition allowed by the ``Landau rules''~\cite{LandauStatPhys} and predicted in Ref.~\onlinecite{Elder04}.
\item In $d=2$ the droplet free energies scale as $F[\psi_{\mathrm{OS}}]\sim\epsilon^{3/2}$
and $F[\psi_{\mathrm{LS}}]\sim\epsilon^{2}$. Lattice structure droplets
are therefore preferred when $\epsilon\ll1$. The reciprocal lattice
vectors form a single triangle, Fig.~\ref{fig:symmetries}(a). In
direct space, the droplet core has a hexagonal lattice structure with
amplitude $\mathcal{O}(\epsilon)$. This is the droplet we observe
numerically, Fig.~\ref{fig:lattice}(a).
\item In $d=3$ the droplet free energies scale identically in all model
parameters ($\epsilon,\sigma,b,R^{d},k_{0}$). The universal ratio,
\begin{equation}
\frac{F[\psi_{\mathrm{LS}}]}{F[\psi_{\mathrm{OS}}]}=\frac{cn\pi^{4}}{m^{2}},
\end{equation}
then determines the preferred droplet structure. Of the possible lattice structure
droplets shown in Fig.~\ref{fig:symmetries}, bcc is preferred because
it minimizes $n/m^{2}=12/4^{2}$. If the constant $c$ were $1$,
 $F[\psi_{\mathrm{LS}}]$ would be 73 times greater than $F[\psi_{\mathrm{OS}}]$!
Our numerical efforts to calculate $c$ have failed, suggesting that
the ansatz $\psi_{\mathrm{LS}}$ does not exist in the spinodal limit,
$\epsilon\ll1$. Instead, we always find onion structure droplets
near the spinodal, Fig.~\ref{fig:onion}. At temperatures
away from both the spinodal and liquid/solid coexistence, the droplet
structure depends on model details. For the clump model, we find
droplets with an icosahedral core, Fig.~\ref{fig:lattice}(b).
\item In $d\geqslant4$ the onion structure droplets have lower free energy,
$F[\psi_{\mathrm{OS}}]\sim\epsilon^{3/2}$, than the lattice structure
ones, $F[\psi_{\mathrm{LS}}]\sim\epsilon^{3-d/2}$, when $\epsilon\ll1$.
Our theory of nucleation to a modulated phase (frequency $k_{0}>0$)
is qualitatively different than classical results for the ferromagnet
($k_{0}=0$)~\cite{Unger84}. In ferromagnetic nucleation, universality
breaks down at $d\geq6$, the upper critical dimension of a $\phi^{3}$
theory~\cite{Muratov04}. In contrast, onion structure droplets have
no such critical dimension.
\end{itemize}

\section{Conclusions\label{sec:conclusions}}

We have demonstrated a new type of nucleating droplet to describe nucleation in supercooled liquids with effective long-range interactions. This onion structure droplet is spherically symmetric and modulated in the radial direction. In two dimensions, droplets with a hexagonal lattice structure, Fig.~\ref{fig:lattice}(a), have a lower free energy barrier than onion structure ones. In three dimensions, in contrast, our scaling arguments and numerical results indicate that onion structure droplets, Fig.~\ref{fig:onion}, are universally favored for a general class of DFT models when quenched near the spinodal.

For shallower quenches the structure of the nucleating droplet depends sensitively on model details. In our numerical study of the clump model away from the spinodal we find a droplet with icosahedral symmetry to be the lowest free energy saddle point solution, Fig.~\ref{fig:lattice}(b). We find that, in its initial growth, this icosahedral droplet develops onion structure modulations away from the core. In studies of atomistic crystallization, a variety of non-universal droplet structures have been observed~\cite{Wolde95,Shen96a,Wolde99,Cherne04,Wang07}. For such systems, DFT models of nucleation near the spinodal may be inapplicable.

\begin{acknowledgments}
K.B.\ was supported by the LANL/LDRD program under the auspices of the DOE NNSA, contract number DE-AC52-06NA25396. W.K.\ was supported by the DOE BES under grant number DE-FG02-95ER14498.
\end{acknowledgments}

\bibliographystyle{apsrev4-1}
\bibliography{/Users/kbarros/Dropbox/Pdfs/bibtex/journals,/Users/kbarros/Dropbox/Pdfs/bibtex/materials/0materials}

\end{document}